\documentstyle[preprint,aps,epsf,titlepage]{revtex}
\begin{document}
\title{Bose Condensation Without Broken Symmetries}
\author{\it Andrei E. Ruckenstein}
\address{Department of Physics and Astronomy, Rutgers University,
136 Frelinghuysen Road, Piscataway, NJ-08854-0849, USA 
\\e-mail: andreir@physics.rutgers.edu}
\date{\today}
\maketitle

%
%     Abstract
%
\begin{abstract}
This paper considers the issue of Bose-Einstein condensation in a
weakly interacting Bose gas with a fixed total
number of particles. We use an old current algebra
formulation of non-relativistic many body systems due to Dashen and Sharp
to show that, at sufficiently low temperatures, a gas of weakly
interacting Bosons displays Off-diagonal Long Range Order in the
sense introduced by Penrose and Onsager. Even though this formulation is
somewhat cumbersome it may demystify many of the
standard results in the field for those uncomfortable with the
conventional broken symmetry based approaches. All the physics presented
here is well understood but as far as we know this perspective, although
dating from the 60's and 70's, has not appeared in the literature. We have
attempted to make the presentation as self-contained as possible in the
hope that it will be accessible to the many students interested in the
field. 
\\ Pacs numbers: 03.75.Fi 05.30.Jp
\end{abstract}
\newpage

\section{Introduction}

The upsurge of activity in recent years in the area of Bose
Einstein Condensation (BEC) in dilute Bose gases has led to remarkable
experimental advances and innovations unimaginable ten years ago when
the first condensate in trapped alkali atoms was reported~\cite{alkali}.
In parallel with these beautiful experiments a better theoretical
understanding is also emerging: These systems are the first experimental
realization of the weakly interacting Bose condensed gas studied long ago
by Bogoliubov, Beliaev, Lee and Yang, and Gross and
Pitaevskii~\cite{history}. Moreover, there are issues specific to the
trapped alkali atoms -- the multicomponent structure of the condensate
due to the internal atomic hyperfine structure, the strongly
inhomogeneous nature of the condensate reflecting the strong
inhomogeneity introduced by the trapping potential, the rather small
number of particles involved in some cases (as low as
$10^5$ compared to the typically $10^{23}$ in liquid helium), as well as
the metastability and sometimes instability of the gaseous phase with
respect to self-binding into liquid or solid droplets -- which make these
systems rich and interesting in their own right. In fact, it is probably
fair to say that a first principles theory, especially as
far as dynamical properties are concerned, is not yet fully developed and
tested~\cite{review}.

This paper was motivated by a heated discussion between the
atomic physics and quantum optics communities on one hand and the
condensed matter physics community on the other, concerning the necessity
of using the notions of broken symmetry and order parameters in
describing the physics of BEC in trapped gases. Indeed, with the
exception of variational ground state wavefunction approaches and
computer simulations, all quantitative many-body treatments of BEC rely on
the notion of a broken symmetry in which the ground state of the system
breaks a global symmetry of the original Hamiltonian~\cite{brokensym}. In
the particular case of BEC this is the global $U(1)$ gauge rotation
generated by $\hat U (\theta) = exp
\left (-i\theta \hat N\right )$ associated with the conservation of the
number of particles, $[\hat H ,\hat N] = 0$. Formally this is
implemented in explicit calculations by using the
so-called  ``symmetry broken ensemble"~\cite{brokensym}
characterized by a density matrix, $\hat \rho$, with the property, $[\hat
\rho , \hat N]\neq 0$, in spite of particle number conservation by the
Hamiltonian. Within this description the BEC displays a finite ``order
parameter",
$\left <\hat \Psi ({\bf r})\right > \neq 0$ ($\left <\hat \Psi ^{\dagger}
({\bf r})\right > \neq 0$), where
$\hat \Psi ({\bf r})$ ($\hat \Psi ^{\dagger} ({\bf r})$) is the Boson
annihilation (creation) operator at position ${\bf r}$, and $\left <
..\right>$ denotes the ground state expectation value (at zero
temperature, $T=0$) or the expectation value in the appropriate
thermodynamic ensemble (for $T\neq 0$). One should imagine that our Bose
gas can extract particles from or inject particles into a sufficiently
large particle reservoir disturbing the strict number conservation; the
Bose condensate would then correspond to a coherent state built from of
superposition of states with different numbers of particles allowed by
the contact with the reservoir. 

It is no surprise that this formal and
unintuitive construction appeared somewhat mysterious to many
unfamiliar with superfluid liquid Helium
physics who felt uneasy in applying these
ideas to an isolated system with a fixed number of particles such as the
trapped gases. This led to a number of reformulations and
generalizations of the Gross-Pitaevskii and Bogoliubov-de Gennes
equations which respect the $U(1)$ symmetry~\cite{fixedN}.
These schemes are more awkward than the conventional
many-body techniques, less amenable to analytical approximations and more
difficult to extend to finite
temperatures and systems far from equilibrium. Here we introduce the
reader to yet another
$U(1)$-symmetric approach based on the current algebra formulation of
non-relativistic many-body systems initiated in the 60's by Dashen and
Sharp~\cite{dashen}. As all other particle conserving schemes this also
has its unsettling features (like, as shown below, the appearance of
singular operators); nevertheless, as the approach only involves
operators which commute with the total number of particles, it can be
studied with well-known field theory techniques and can be easily
extended to treat finite temperatures and non-equilibrium situations.
We note that the current algebra description can be interpreted as the
operator version of the
$U(1)$-symmetric functional integral representation due to
Popov~\cite{popov}. Different aspects of Bose condensation using the
Dashen-Sharp current algebra formulation have been discussed in the
past~\cite{others} but, as far as we know ours is (i) the first
discussion of inhomogeneous systems and (ii) the first computation of
single-particle correlation functions.

Historically, the description of BEC in systems with a {\em fixed number
of particles} follows from the
realization~\cite{penons} that, in the thermodynamic limit (with
$N\rightarrow \infty$ with the volume
$\Omega
\rightarrow \infty$ with $n=N/\Omega$ fixed), the Bose condensed state
displays long-ranged correlations in the single particle density matrix,
$\rho( {\bf{r}} ;
{\bf{r'}}) =  \left < \Psi ^{\dagger} ({\bf{r}}) \Psi ({\bf{r'}})\right
>$, where
$\Psi ^{\dagger} ({\bf{r}}) (\Psi ({\bf{r}}))$ is the creation
(annihilation) operator for a boson at site ${\bf r} = (x,y,z)$. More
precisely, as
$\rho ({\bf r}, {\bf r'})$ is hermitian with respect to the ${\bf r}$ and
$\bf{r'}$ indices, it can be expanded in terms of its complex
eigenfunctions, $\phi _{\alpha} (\bf{r})$ and real eigenvalues,
$\Lambda_{\alpha}$ as,
\begin{equation}
\label{ODLRO}
\rho ( {\bf{r}} ; {\bf{r'}}) ~=~ 
\left < \Psi ^{\dagger} ({\bf{r}}) \Psi ({\bf{r'}})\right > ~=~\sum
_{\alpha} 
\Lambda _{\alpha} \phi _{\alpha}  ^* (\bf{r}) \phi _{\alpha}
(\bf{r'}).
\end{equation}
BEC in the noninteracting gas is signaled by
a macroscopic eigenvalue associated with the zero momentum state,
$\Lambda _{\bf{k} =0} = n^{(0)} _0 (T)$, where in this case $\alpha$
labels momentum eigenstates $\phi _{\bf{k}} ({\bf{r}}) = [exp
(i\bf{k}\cdot \bf{r})]/\sqrt{\Omega}$, $n^{(0)} _0 (T)$ is the condensate
density of the noninteracting gas at temperature $T$ and $\Omega$ is the
volume of the system. BEC displays long range correlations in the sense
that
$lim_{|\bf{r} -\bf{r'}|\rightarrow\infty} \rho ( {\bf{r}} ; {\bf{r'}})
= n^{(0)} _0 (T) \phi _{\bf{k} =0}  ^* (\bf{r}) \phi
_{\bf{k}=0} (\bf{r'})$, while the contributions from higher momentum
states oscillate away in the limit. By adiabatic continuity,
condensation in the interacting Bose gas is defined by the presence of
{\em one} eigenfunction of the single particle density matrix, $\phi _0
(\bf{r} )$~\cite{fragmentation}, with a macroscopic eigenvalue, $\Lambda
_0 = n_0$. It's not hard to show that, at $T=0$, $\Phi ({\bf{r}}) \equiv
\sqrt{n_0} \phi _0 (\bf{r})$, often referred to as ``the macroscopic
wave-function" satisfies a non-linear Schr\"{o}dinger equation,
the time independent version of the Gross-Pitaevskii 
equation~\cite{review}. In this case the system is said to display
Off-diagonal Long Range Order (ODLRO) but $\left <\Psi ({\bf
r})\right> =0$ and thus, strictly speaking, the symmetry
remains unbroken~\cite{comment} 

Below we use the current algebra approach in a Bose gas at zero
temperature to calculate the low
energy excitation spectrum and to check for the presence of BEC in the
ground state, in the sense of Penrose and Onsager~\cite{penons}.

\section{The Current Algebra Approach}

The discussion that follows is based in its entirety on the work of
Dashen and Sharp~\cite{dashen}. The
idea is to represent the Hamiltonian and all operators of the theory in
terms of the density,
$\hat{n} ( {\bf{r}}) = \Psi ^{\dagger} ({\bf{r}})
\Psi ({\bf{r}})$ and current $\vec{\hat{\bf{g}}} ({\bf{r}}) = (-i /2M)
[\Psi ^{\dagger} ({\bf{r}})\vec{\nabla} \Psi ({\bf{r}}) - \vec{\nabla}
\Psi ^{\dagger} ({\bf{r}}) \Psi ({\bf{r}})]$ operators which obey the
current algebra:
\begin{eqnarray}
\label{curalg1}
[\hat{n} ( {\bf{r}}),\hat{n} ( {\bf{r'}})]&=&0 \\
\label{curalg2}
\left[\hat{n} ( {\bf{r}}), \vec{\hat{\bf{g}}} ({\bf{r'}}) \right] &=&
-\frac{i}{M}
\vec{\nabla}_{\bf{r}} (\delta ({\bf{r}}-{\bf{r'}}) {\hat{n}} (
{\bf{r}}))\\
\label{curalg3}
\left[ \hat{\bf{g}}^{\alpha} 
({\bf{r}}), \hat{\bf{g}}^{\beta} ({\bf{r'}}) \right] &=&
-\frac{i}{M}
\left [{\nabla}^{\beta} _{\bf{r}} (\delta ({\bf{r}}-{\bf{r'}})
\hat{\bf{g}}^{\alpha} ({\bf{r}}))~-~
{\nabla}^{\alpha} _{\bf{r'}} (\delta ({\bf{r}}-{\bf{r'}})
\hat{\bf{g}}^{\beta} ({\bf{r'}}))\right ].
\end{eqnarray} 
(Hereafter $\alpha ,
\beta = x, y ,z$ index vector components,
$\hbar =1$, and the hat will differentiate operators from
classical fields whenever ambiguities can arise.)

The main step of this formulation is to rewrite
the Bose Hamiltonian in terms of $\hat{n} ( {\bf{r}})$ and
$\vec{\hat{\bf{g}}} ({\bf{r}})$ by using unity in the form, ${\bf{I}} =
\Psi ({\bf{r}})[1/\hat{n} ( {\bf{r}})]
\Psi ^{\dagger} ({\bf{r}})$ ($\hbar =1$)~\cite{projection}:
\begin{eqnarray}
\label{ham}
H&=&\int
\left\{\frac{1}{2M}
\left(\vec{\nabla} \Psi ^{\dagger} ({\bf{r}})\right )
\left [\Psi ({\bf{r}}) \frac{1}{\hat{n} ( {\bf{r}})} \Psi
^{\dagger} ({\bf{r}})\right ]\left( \vec{\nabla} \Psi
({\bf{r}})\right)+\frac{1}{2}
M\omega _0 |\vec{\bf{r}}|^2\hat{n}({\bf{r}})+\frac{g}{2}
\hat{n} ^2 ({\bf{r}})
\right\} d{\bf{r}}\\
&=&\int \left\{\frac{1}{8M} \left[\vec{\nabla} \hat{n} ( {\bf{r}})
-2iM\vec{\hat{\bf{g}}} ({\bf{r}})\right]
\frac{1}{\hat{n} ( {\bf{r}})} \left [\vec{\nabla} \hat{n}( {\bf{r}})~
+~2iM \vec{\hat{\bf{g}}} ({\bf{r}})\right] +\frac{1}{2}
M\omega _0 |\vec{\bf{r}}|^2\hat{n}({\bf{r}})+ \frac{g}{2}
\hat{n} ^2 ({\bf{r}})
\right\} d{\bf{r}}\nonumber,
\end{eqnarray}
where for simplicity the interparticle potential was replaced by the
s-wave pseudopotential, $v({\bf{r}} - {\bf{r'}}) = g \delta
({\bf{r}}-{\bf{r'}})$, with the strength, $g=4\pi a/M$, written in terms
of the s-wave scattering length, $a$; and use was made of the identities,
$(\vec{\nabla} \Psi ^{\dagger} ({\bf{r}}) \Psi ({\bf{r}})) =
(\vec{\nabla} \hat{n} ( {\bf{r}}) -2iM\vec{{\hat{\bf{g}}}} ({\bf{r}})
)/2$ and
$(\Psi ^{\dagger} ({\bf{r}})\vec{\nabla}
\Psi ({\bf{r}})) = (\vec{\nabla} \hat{n} ( {\bf{r}}) +
2iM\vec{{\hat{\bf{g}}}}
({\bf{r}}) )/2$. We have included an external
harmonic potential (with frequency $\omega _0$) which traps
the particles in a finite region of space.

Representations of the current algebra (\ref{curalg1}), (\ref{curalg2})
and (\ref{curalg3}) have been discussed extensively in the
70's~\cite{curalg}. It was shown that in an irreducible $N$-particle
representation of the current algebra matrix elements of operators such as
$(\vec{\nabla} \hat{n} ( {\bf{r}}) -2iM\vec{{\hat{\bf{g}}}} ({\bf{r}}))$
and
$(\vec{\nabla} \hat{n} ( {\bf{r}}) +2iM\vec{{\hat{\bf{g}}}} ({\bf{r}}) )$
are proportional to $\hat{n} ( {\bf{r}})$, and thus the singular operator,
$1/\hat{n} ( {\bf{r}})$, in (\ref{ham}), disappears in physical
matrix elements. Below we
will leave aside all rigor and manipulate expressions involving 
$1/\hat{n} ( {\bf{r}})$ formally with the expectation that, even if
intermediate states of some of the calculations are ill-defined, the final
answer is physically meaningful.

\subsection{The Excitation Spectrum}

To obtain the low lying excitation spectrum we use a mean field
approximation and expand the Hamiltonian (\ref{ham}) to quadratic order in
fluctuations of the density and current around their values in the ground
state. Below we only describe the more general case of the nonuniform
system and extract the homogeneous gas results as a special limit.

In the presence of the harmonic potential the ground state, 
$\left|\Omega \right > $, is characterized by an inhomogeneous particle
density, 
$\left < \Omega | \hat{n} ({\bf{r}}) |\Omega \right > = n_G ({\bf{r}})$,
and a (particle) current density, $\left <\Omega |\vec {\hat{\bf{g}}}
({\bf{r}})|\Omega \right > =
\vec{\bf g} _G ({\bf r})$, to be determined by minimizing the mean-field
energy functional,
\begin{equation}
\label{meanfield}
E_{GS} [n_G ({\bf r}), \vec{g} _G ({\bf r})] = \int \left\{
\frac{[(\vec{\nabla} n_G ( {\bf{r}}))^2 + 4M^2 (\vec{{\bf g}}_G 
({\bf{r}}))^2 ]}
{8Mn_G ({\bf r})}
 +\frac{1}{2}
M\omega _0 |\vec{\bf{r}}|^2 n_G ({\bf{r}})+ \frac{g}{2}
n_G ^2 ({\bf{r}})
\right\} d{\bf{r}},
\end{equation}
subject to the fixed particle number constraint, $\int d{\bf r} n_G ({\bf
r}) = N$. As usual this constraint is enforced by
adding a chemical potential like term to (\ref{meanfield}), $E_{GS} [n_G
({\bf r}), \vec{g} _G ({\bf r})]\rightarrow E_{GS} [n_G ({\bf r}),
\vec{g} _G ({\bf r})] -\mu \int d{\bf r} n_G ({\bf r})$. The resulting
ground state carries no current, $\vec{\bf g} _G ({\bf r})=0$, and a
nonuniform density  satisfying,
\begin{equation}
\label{gpeq} 
\frac{(\vec{\nabla} _{\bf r} n_G ({\bf r}))^2}{8M n_G ^2
({\bf r})}~-~\frac{\nabla _{\bf r} ^2 n_G ({\bf r})}{4M n_G ({\bf r})}~ 
+ ~\left (\frac{1}{2} M\omega _0 ^2 |\vec{\bf r}|^2 -\mu
\right) ~+~ g n_G ({\bf r})=~0
\end{equation}

The low lying, large length scale excitations of the system
are described by the effective Hamiltonian, $H_X$, obtained from
(\ref{ham}) by making the replacement $\hat
{n} ({\bf{r}}) = n_G ({\bf{r}}) + \hat\eta ({\bf{r}})$ and keeping the
terms leading (i.e., quadratic) order in the excitation operators,
$\hat \eta ({\bf{r}})$ and
$\vec {\hat {\bf{g}}} ({\bf{r}})$:
\begin{eqnarray}
\label{hamx}
\hat H _X&=&\int d{\bf r} \left \{ \frac{1}{8Mn_G ({\bf r})} \left [
\left (
\vec{\nabla}
_{\bf r} \hat \eta ({\bf r}) \right ) ^2+4M^2 \left ({\vec{\hat {\bf g}}}
({\bf r})\right ) ^2\right ]+ \left [\frac{1}{8M n_G ({\bf
r})} \left ( \frac{\vec{\nabla} _{\bf r} n_G ({\bf r})}{n_G ({\bf
r})}\right ) ^2 +\frac{g}{2} \right ]\hat \eta ^ 2({\bf r}) \right \}
\nonumber\\
&-&\int d{\bf r}\frac{\vec{\nabla} _{\bf r} n_G ({\bf r})}{8M n_G ^2 ({\bf
r})}\cdot
\left [
\hat \eta ({\bf r})
\left ( \vec{\nabla} _{\bf r}\hat
\eta ({\bf r}) +2iM\vec{\hat {\bf g}} ({\bf r})\right )+\left (
\vec{\nabla} _{\bf r}\hat
\eta ({\bf r}) -2iM{\vec{\hat {\bf g}}} ({\bf r})\right )\hat \eta ({\bf
r})\right ].
\end{eqnarray}
For the purpose of obtaining the linear excitations, the current algebra
(\ref{curalg1}), (\ref{curalg2}) and (\ref{curalg3}) is replaced by the
(linearized) approximation, 
\begin{eqnarray}
\label{curalginh}
[\hat{n} ( {\bf{r}}),\hat{n} ( {\bf{r'}})]&=&0 ; 
\left[ \hat{\bf{g}}^{\alpha} ({\bf{r}}), \hat{\bf{g}}^{\beta}
({\bf{r'}}) \right] \approx 0\\
\left[\hat{\eta} ( {\bf{r}}), \vec{\hat{\bf{g}}} ({\bf{r'}}) \right]
&\approx& -\frac{i}{M}
\vec{\nabla}_{\bf{r}} (\delta ({\bf{r}}-{\bf{r'}}) n_G
({\bf{r}})).
\end{eqnarray}

With these simplifications the equations of motion for the excitation
operators, read:
\begin{eqnarray}
\label{conteq}
\frac{\partial}{\partial t} \hat \eta ({\bf r}, t) ~&=&~-\vec{\nabla}
_{\bf r} 
\cdot \vec{\hat {\bf g}} ({\bf r},t)\\
\label{cureq}
\frac{\partial}{\partial t} \hat {\vec{\bf g}} ({\bf r},
t)~&=&~\frac{1}{4M^2} n_G ({\bf r}) \vec{\nabla} _{\bf r} \left
\{\nabla _{\bf r} ^{\alpha} \left ( \frac{\nabla ^{\alpha} _{\bf r}
\hat
\eta ({\bf r}, t)}{n_G ({\bf r})}\right )+\hat \eta ({\bf r},t)
\left [\nabla _{\bf r} ^2 \left (1/{n_G
({\bf r})} \right )-4 \left (\vec{\nabla} _{\bf r} \left (1/{\sqrt
{n_G ({\bf r})}} \right )\right ) ^2\right ]\right \}\nonumber\\
&-&{\frac{g}{M} n_G ({\bf r}) \vec
{\nabla} _{\bf r}} 
\hat
\eta ({\bf r}, t)
\end{eqnarray}
where summation over repeated indices is implied.
Naturally the density fluctuations satisfy
a continuity equation (\ref{conteq}). 

For illustration consider the
Thomas-Fermi (TF) limit~\cite{TF} in which one ignores the gradient
terms in (\ref{gpeq}) and in the curly bracket in (\ref{cureq}).
In that case,
$g n_G ({\bf r}) \approx  M\omega _0 ^2 (R^2 -|\vec {\bf r}|^2 )/2$, 
and Equations (\ref{conteq}) and (\ref{cureq}) combine to give,
\begin{equation}
\frac{\partial ^2}{\partial t^2} \hat \eta ({\bf r}, t) ~=~\frac{1}{2}
\omega _0 ^2 \nabla _{\bf r} ^{\alpha} \left [ (R^2 - |\vec {\bf r}|^2)
\nabla _{\bf r} ^{\alpha} \hat \eta ({\bf r} ,t)\right].
\end{equation}
($R$ is the spatial extent of the condensate which, for a spherical trap
with N particles, is given by $R=a_{ho} \left (15 Na/a_{ho}\right
)^{1/5}$ in terms of the harmonic oscillator length,
$a_{ho}$ and the s-wave scattering length, $a$.) This is precisely the
equation first discussed by Stringari~\cite{string} and extensively
studied since for a variety of trap geometries~\cite{review}.

It is clear that to obtain the results of the uniform
system one must go beyond the TF approximation. This can be done
either directly from equations (\ref{gpeq}), (\ref{conteq}) and
(\ref{cureq}) or by making contact with the conventional
non-conserving Bogoliubov approach~\cite{fetter}. We will take the latter
route.

\subsubsection{Correspondence to the Bogoliubov Approach}

We proceed by first noticing that the identity, $\vec{\nabla}_{\bf{r}}
(\delta ({\bf{r}}-{\bf{r'}}) n_G ({\bf{r}}))= - n_G ({\bf r'})\nabla
_{\bf r'}
\delta ({\bf r} -{\bf r'})$ suggests rewriting the current operator as
$\vec {\hat {\bf g}} ({\bf r}) = n_G ({\bf r}) \vec \nabla _{\bf r} 
\hat \varphi ({\bf r})/M$, where the operator $\hat \varphi ({\bf r})$ is
the canonically conjugate momentum density to $\hat \eta ({\bf r})$:
$[\hat
\eta ({\bf r}),\hat \varphi ({\bf r'})] = i\delta ({\bf r} -{\bf r'})$ and
$[\hat \varphi ({\bf r}),\hat \varphi ({\bf r'})] = 0$. As in the
example of the simple harmonic oscillator we can then make linear
combinations of ``coordinates" ($\hat \eta ({\bf r})$) and ``momenta" 
($\hat \varphi ({\bf r})$) to
construct creation ($\hat b^{\dagger} ({\bf r})$) and annihilation
($\hat b ({\bf r})$) operators. Here we choose,
\begin{eqnarray}
\label{boscre}
\sqrt {n_G ({\bf r})} {\hat b} ^{\dagger}~ ({\bf r}) &=& \frac{1}{2}
\left ( \hat \eta ({\bf r}) - 2i n_G ({\bf r}) 
\hat \varphi ({\bf r})\right)\\
\label{bosan}
\sqrt {n_G ({\bf r})} {\hat b} ({\bf r}) &=& \frac{1}{2} \left ( \hat
\eta ({\bf r}) + 2i n_G ({\bf r}) \hat \varphi ({\bf r})\right )\\
\left [\hat b ({\bf r}) , \hat b ^{\dagger} ({\bf r'}) \right ] &=& \delta
({\bf r} - {\bf r'}); \left [\hat b ({\bf r}) , \hat b ({\bf r'}) \right
]=\left [\hat b ^{\dagger} ({\bf r}) , \hat b ^{\dagger} ({\bf r'}) \right
]=0.
\end{eqnarray}

It is not hard to see that, in terms of the
new canonical Bose fields (\ref{hamx}) 
becomes,
\begin{equation}
\label{bog}
\hat H _X \approx \int d{\bf r} \left [ \frac{1}{2M} \left (\vec{\nabla}
_{\bf r} \hat b^{\dagger} ({\bf r})\right ) \cdot \left (\vec{\nabla}
_{\bf r} \hat b ({\bf r})\right )~+~\frac{g}{2} {n_G ({\bf r})} \left
(\hat b ({\bf r}) +\hat b ^{\dagger} ({\bf r})\right)^2\right],
\end{equation}
where the only term in (\ref{hamx}) incorrectly reproduced in
(\ref{bog}) are those involving the $\nabla _{\bf r} n_G ({\bf r})$ terms
proportional to $\hat {\eta} ^2 ({\bf r})$. To recover these
terms and to eliminate unwanted terms proportional to $\hat{\varphi} ^2
({\bf r})$ requires going beyond the linear approximations (\ref{boscre})
and ({\ref{bosan}) in the representation of
${\hat b} ({\bf r})$  and ${\hat b} ^{\dagger} ({\bf r})$ in terms of
$\hat {\eta} ({\bf r})$ and 
$\hat {\varphi} ({\bf r})$. (Note that all the problematic higher order
terms involve gradients of $n_G ({\bf r})$ and thus vanish for a
uniform system.)

The Hamiltonian (\ref{bog}) is identical to that
derived by Bogoliubov in the presence of a finite condensate order
parameter~\cite{fetter}; for the case of a fixed number of particles the
order parameter vanishes and
$H_X$ should be interpreted not as a quasi-particle Hamiltonian but as
the Hamiltonian describing the low lying {\em density and current
excitations} of the system at a fixed total particle number. It is then
no surprise that (\ref{bog}) does not conserve the number of bosons,
$\int d{\bf r}\hat b ^{\dagger} ({\bf r})
\hat b ({\bf r})$.

The quadratic form $H_X$ is easily diagonalized by the
Bogoliubov transformation, $\hat b ({\bf r}) = \sum _n \left (
u_n ({\bf r}) \hat \beta _n ({\bf r}) - v_n ^* ({\bf r}) \hat \beta _n
^{\dagger} ({\bf r})\right )$ and $\hat b ^{\dagger} ({\bf r}) = \sum _n
\left ( u_n ^*({\bf r}) \hat \beta _n ^{\dagger} ({\bf r}) - v_n ({\bf
r}) \hat \beta _n ({\bf r})\right )$,
with the functions $u_n ({\bf r})$ and $v_n ({\bf r})$ satisfying,
\begin{eqnarray}
\left (-\frac{\nabla _{\bf r} ^2}{2M}~+~gn_G ({\bf r}) \right ) u_n ({\bf
r})~-~gn_G ({\bf r}) v_n ({\bf r}) ~&=&~ E_n u_n ({\bf r})\\
\left (-\frac{\nabla _{\bf r} ^2}{2M}~+~gn_G ({\bf r}) \right ) v_n ({\bf
r})~-~gn_G ({\bf r}) u_n ({\bf r}) ~&=&~ -E_n v_n ({\bf r})
\end{eqnarray}
and the orthonormality condition, $\int d{\bf r} \left (
u_n ^* ({\bf r}) u_{n'} ({\bf r})- v_n^ * ({\bf r}) v_{n'}({\bf r})
\right ) = \delta _{n,n'}$.

For a uniform system, $u_{\bf k} ({\bf r}) = u_{\bf k} e^{i{\bf
k}\cdot{\bf r}}$, $v_{\bf k} ({\bf r}) = v_{\bf k} e^{i{\bf
k}\cdot{\bf r}}$, $n_G =n=N/\Omega$, and these equations immediately lead
to the well known Bogoliubov results~\cite{fetter},
$u _{\bf k} =\frac{1}{\sqrt 2}
\left [E^{-1} _{\bf k} \left (\epsilon _{\bf k}
+g n\right ) +1\right]^{\frac{1}{2}}$ and $v_{\bf k} =-\frac{1}{\sqrt 2}
\left [E^{-1} _{\bf k} \left (\epsilon _{\bf k}
+g n\right ) -1\right]^{\frac{1}{2}}$, where
$E_{\bf k} =  \sqrt{\epsilon _{\bf k} (\epsilon _{\bf k} +2gn)}$
($\epsilon _{\bf k} = |{\bf k}|^2 /2M$) is the Bogoliubov quasiparticle
energy.

\subsection{Off-Diagonal Long Range Order}

To check for the occurrence of BEC in the sense of ODLRO is more involved
as we need to calculate the single particle density matrix,
$\rho ({\bf{r}}+{\bf{i}}x ,{\bf{r}}) = \left <\hat\Lambda
({\bf{r}}+{\bf{i}} x;{\bf{r}}) \right>$ where
$\hat\Lambda ({\bf{r}}+{\bf{i}} x;{\bf{r}}) = \Psi ^{\dagger}
({\bf{r}}+{\bf{i}} x) \Psi({\bf{r}})$. At first sight it is hard to see
how one might do this computation in an approach in which the basic
variables are number and current densities. The solution can be found
in the old work of Grodnik and Sharp~\cite{dashen}.

Without loss of generality we take
the separation vector between the two points,
${\bf{r}}$ and
${\bf{r'}}$, in $\rho ({\bf{r}},{\bf{r'}})$, along the $x$-axis.
We then proceed by considering the differential
equation,
\begin{eqnarray}
\label{corrfnc}
\nabla _x  \hat\Lambda ({\bf{r}}+{\bf{i}} x;{\bf{r}})~& = &~\nabla _x \Psi
^{\dagger} ({\bf{r}}+{\bf{i}} x) \Psi({\bf{r}})~ =~
\nabla _x \Psi ^{\dagger} ({\bf{r}}+{\bf{i}} x) \Psi
({\bf{r}} +{\bf{i}} x) \frac{1}{\hat{n}( {\bf{r}}+{\bf{i}} x)}
\Psi ^{\dagger} ({\bf{r}} +{\bf{i}} x) \Psi({\bf{r}})\nonumber\\
&=& \frac{1}{2} \left [ \nabla _x \hat{n}( {\bf{r}} +{\bf{i}} x) - 2iM
\hat {\bf g} _x ({\bf r} +{\bf i}x)\right ]\frac{1}{\hat{n}( {\bf r} +{\bf
i} x)}\hat\Lambda ({\bf{r}}+{\bf{i}} x;{\bf{r}}),
 \end{eqnarray}
where in the last two steps we used the resolution of the identity,
${\bf{I}} = \Psi ({\bf{r}})[1/\hat{n} ( {\bf{r}})]
\Psi ^{\dagger} ({\bf{r}})$. Together with the initial condition,
$\hat\Lambda ({\bf{r}}+{\bf{i}} x;{\bf{r}})|_{x=0} = \hat{n}( {\bf{r}})$,
and the linearized approximation, $\left [\nabla _{x'}
\hat{n}( {\bf{r}} +{\bf{i}} x') - 2iM \hat {\bf g} _x ({\bf r} +{\bf
i}x')\right ]\frac{1}{2\hat{n}( {\bf r} +{\bf i} x')}\approx
\nabla _{x'} \left (\hat b ^{\dagger} ({\bf r}+{\bf i}x') /\sqrt{n_G
({\bf r}+{\bf i}x')}\right)$, Eq. (\ref{corrfnc}) leads to a
``x-ordered" exponential which allows us to write down the general
formula for the spatial correlation function:
\begin{eqnarray}
\label{rhox}
\rho ({\bf{r}} +{\bf i} x , {\bf r}) & \equiv & \tilde \rho
(|x|)=\left< T_{x'} 
\left\{ exp {\int _{0}
^{x} dx'
\frac{1}{2} \left[ \nabla _{x'} \hat{n}( {\bf{r}} +{\bf{i}} x') - 2iM
\hat {\bf g} _x ({\bf r} +{\bf i}x')\right]\frac{1}{\hat{n}( {\bf r}
+{\bf i} x')}}\right\}
\hat{n} ( {\bf{r}}) \right>\nonumber\\
& \approx &\sqrt {n_G({\bf r}+{\bf i}x) n_G ({\bf r})} \left < 
\left[exp \left ( \frac{\hat b ^{\dagger} ({\bf r}+{\bf i}x)}{\sqrt{n_G
({\bf r}+{\bf i}x)}} - \frac{\hat b ^{\dagger} ({\bf r})}{\sqrt{n_G ({\bf
r})}}\right ) \right ] 
\left [1+\left (\frac{\hat{b} ({\bf r}) +\hat b ^{\dagger} ({\bf
r})}{\sqrt {n_G ({\bf r})}}\right )\right ] \right >.
\end{eqnarray}

The explicit computation of (\ref{rhox}) is then carried out by
transforming to Bogoliubov quasiparticles and using the
``disentangling" Baker-Hausdorff formula, $exp (\hat A +\hat B) = exp
-\frac{1}{2} [\hat A ,\hat B] exp
\hat A exp \hat B$ (where $[[\hat A ,\hat B], \hat A] = [[\hat A ,\hat
B], \hat B]=0$) together with the condition
that $\hat \beta ({\bf r})$ annihilates the ground state, $\hat
\beta ({\bf r}) |\Omega\rangle = 0$.
The resulting expression, now written for an arbitrary separation vector,
${\bf R}$, is expressed in terms of the Bogoliubov amplitudes,$u_n ({\bf
r})$ and
$v_n ({\bf r})$, as:
\newpage
\begin{eqnarray}
\label{rho}
\rho (|{\bf R}|) &=& \left\{ exp\left [-\frac{1}{2} \sum _n \left 
(\frac{u_n ^* ({\bf r}+{\bf R})}{\sqrt{n_G ({\bf r}+{\bf R})}} -
\frac{u_n ^* ({\bf r})}{\sqrt{n_G ({\bf r})}}\right )
\left (\frac{v_n ({\bf r}+{\bf R})}{\sqrt{n_G ({\bf r}+{\bf R})}} -
\frac{v_n ({\bf r})}{\sqrt{n_G ({\bf r})}}\right )\right ]\right
\}\times\nonumber\\
&\times& \left [1-\sum _n \left 
(\frac{u_n ^* ({\bf r}) -v_n ^* ({\bf r})}{\sqrt{n_G ({\bf r})}}\right )
\left 
(\frac{v_n ({\bf r}+{\bf R})}{\sqrt{n_G ({\bf r}+{\bf R})}} -
\frac{v_n ({\bf r})}{\sqrt{n_G ({\bf r})}}\right )\right ]\sqrt{n_G
({\bf r}+{\bf R}) n_G ({\bf r})}. 
\end{eqnarray}

A similar calculation can be done for time dependent single particle 
correlation functions. For example, the local, time-dependent
single-particle density matrix,
$\rho ({\bf r}, t; {\bf r}, 0) = \left <\hat\Lambda
({\bf{r}},t;{\bf{r}},0) \right>$, where
$\hat\Lambda ({\bf{r}},t;{\bf{r}},0) = \Psi ^{\dagger}
({\bf{r}},t) \Psi({\bf{r}})$, is given by:
\begin{eqnarray}
\label{rhot}
\rho ({\bf{r}},t;{\bf r},0) &=&\left
<T_{\tau} 
\left \{ exp {\int _{0}
^{t} d\tau
\frac{1}{2} \left [ \partial _{\tau} \hat{n}(
{\bf{r}},\tau)+\left (\partial _{\tau} \Psi ^{\dagger} ({\bf
r}, \tau)
\Psi ({\bf r},\tau) - \Psi ^{\dagger} ({\bf r},\tau)\partial _{\tau} \Psi
({\bf r},\tau)\right )\right
]\frac{1}{\hat{n}({\bf r},\tau)}} \right\}
\hat{n}( {\bf{r}})\right >\nonumber\\
& \approx&n_G ({\bf r}) e^{i\mu t} \left
<T_{\tau} 
\left [ exp {\int _{0}
^{t} d\tau
\left (\frac{1}{2} \frac{\partial
_{\tau}
\hat{\eta}( {\bf{r}},\tau)}{n_G ({\bf r})}-i\partial _{\tau} \hat \varphi
({\bf r},\tau)\right )}\right ] \left [1+\frac{\hat \eta ({\bf r})}{n_G
({\bf r})}\right ]\right >\\
&=&n_G({\bf r}) e^{i\mu t}
\left < \left[exp \left ( \frac{\hat b ^{\dagger} ({\bf r},t) -
\hat b ^{\dagger} ({\bf r})}{\sqrt{n_G ({\bf r})}}\right ) \right ] 
\left [1+\left (\frac{\hat{b} ({\bf r}) +\hat b ^{\dagger} ({\bf
r})}{\sqrt {n_G ({\bf r})}}\right )\right ] \right >\nonumber.
\end{eqnarray}
The last two equalities in (\ref{rhot}) involve the linear
approximation defined by (\ref{boscre}) and (\ref{bosan}). As with
(\ref{rho}), the final answer is expressible in terms of
Bogoliubov amplitudes,
$u_n ({\bf r})$ and 
$v_n ({\bf r})$ as:
\begin{eqnarray}
\label{rho'}
\rho ({\bf r}, t;{\bf r},0)&=&n_G ({\bf r}) e^{i\mu t} \left \{ exp\left
[-\sum _n \frac{u_n ^* ({\bf r}) v_n ({\bf r})}{n_G ({\bf r})} \left
(1-cos E_n t\right )\right ]\right\}\times\nonumber\\
&\times& \left [1+\sum _n \frac{v_n ({\bf r}) \left (u_n ^* ({\bf r}) -
v_n ^* ({\bf r})\right )}{n_G ({\bf r})}\left (1-e^{-iE_n t} \right
)\right ]
\end{eqnarray}

It is easy to see that in the homogeneous
case (\ref{rho}) and (\ref{rhot}) become:
\begin{eqnarray}
\label{assymp}
lim _{|{\bf R}| \rightarrow \infty} \rho (|{\bf R}|)& =& e^{-i\mu t}
lim _{t \rightarrow \infty} \rho ({\bf r},t;{\bf r},0)\nonumber\\
&=&n\left [ exp \left (-\frac{1}{2\Omega} \sum _{\bf k}
\frac{g}{E_{\bf k}}\right )\right ]\left [1+\frac{1}{2\Omega n}\sum
_{\bf k}
\left (1-\frac{\epsilon _{\bf k}}{E_{\bf k}}\right )\right ].
\end{eqnarray}
A number of features of (\ref{assymp}) are worth noting: (i) as expected
the condensate density is equal to the total density in the
noninteracting limit; (ii) expanding to leading order in
$g$ gives the same result for the depletion of the condensate as
calculated from the conventional Bogoliubov approach~\cite{fetter} (note
that the  ultraviolet cutoff required due to the linear
approximation cancels to leading nontrivial order in $g$); (iii) in a
system with a fixed number of particles the phase of the condensate
contribution precesses uniformly at a rate determined by the chemical
potential, $\mu$, in agreement with the Josephson
relations~\cite{josephson}; (iv)
finally, as expected on general grounds a
$T=0$ condensate occurs in two and three dimensions, whereas in
one-dimension the single-particle correlation function decays
algebraically both in space and in time (due to the infrared logarithmic
divergence in the exponent).
An explicit analysis of the nonuniform Bose gas, including a discussion of
the meaning of ODLRO in finite systems
and a comparison with the results of Ma and Ho~\cite{ho}, is left to a
future publication.

\section{Conclusions}

We have considered the Bose gas in a general inhomogeneous potential
within the current algebra approach to non-relativistic many-body systems
due to Dashen and Sharp~\cite{dashen}. Not surprisingly, we arrive at the
same physics as that of broken symmetry approaches. The
differences are somewhat subtle: strictly speaking, in the broken
symmetry case the order parameter field displays collapse and revivals due
to fluctuations in the particle number~\cite{fixedN}. As a result, the
Gross-Pitaevskii and Bogoliubov-de Gennes equations are adequate only for
times short compared to the collapse time scale which becomes infinite
only in the thermodynamic limit (in the uniform Bose gas $\tau
_{collapse} \sim
\sqrt N$). 
This effect was understood already in the 50's through
Anderson's classic discussion of broken symmetry in quantum
antiferromagnets~\cite{pwa}, for which, as in Bose condensates, the order
parameter is not a constant of the motion. In all such systems, apart
from the Goldstone modes present as a result of the broken symmetry,
there exist modes -- the ``phase diffusion" mode in the case of BEC -- 
with a frequency which vanishes in the infinite volume limit faster than
that of the lowest Goldstone mode. Our
fixed-$N$ formulation only includes the physics of a fixed $N$ sector
within which the phase precesses at a constant rate given by the chemical
potential. Collapse and revivals can be obtained only by averaging over
systems with different values of $N$, as would be the case if our
condensate was brought into contact with an ideal particle reservoir.

Controlled $U(1)$-symmetric techniques are especially important
in treating small systems and systems far from equilibrium where spurious
dynamics of the condensate in symmetry broken ensembles may confuse
some of the important physics. We
expect that the theoretical frameworks presented here and in \cite{fixedN}
can be used to analyze the feasibility of a number of novel experiments on
phase coherence and non-linear atom-optics of condensates. The remarkably
powerful techniques for manipulating atomic condensates
perfected in recent years suggest experiments difficult to imagine in the
context of He superfluids or superconductors, such as those now
common-place in nonlinear and quantum optics, from four-wave mixing,
parametric amplification, squeezed states, to Quantum Electrodynamics of
cavity Bose condensates.

\acknowledgements
It is a pleasure to dedicate this paper to Martin
Gutzwiller on the occasion of his 75th birthday. We are
grateful to A. Kuklov for innumerable conversations on the topic of phase
coherence in isolated Bose condensates, and E.
Abrahams, D. Villani and especially P. C. Hohenberg for comments on the
manuscript. We would also like to thank the organizers and participants
of the 1999 Workshop on Correlated Systems at the Institut Henri
Poincar\'{e} in Paris, France, and of the 1999 Training Course on
Correlated Systems School on Correlated Systems in Salerno, Italy, where
these ideas were first presented.


\begin{thebibliography}{99}
\bibitem{comment} We note that, by contrast, symmetry breaking also
implies ODLRO in the correlation function (\ref{ODLRO}).
\bibitem{fragmentation} Here we will not consider fragmented
condensates.
\end{thebibliography}
\end{document}